\documentclass[twocolumn,showpacs,aps,prl]{revtex4}
\usepackage{graphicx}
\usepackage{dcolumn}
\usepackage{bm}
\def\vep{{\varepsilon}}

\begin{document}
\title{Comments on ``Observation of Strong Quantum Depletion in
a Gaseous Bose-Einstein Condensate" [cond-mat/0601184]}
\author{Sang-Hoon \surname{Kim}
}
\affiliation{Division of Liberal Arts,
Mokpo National Maritime University, Mokpo 530-729, Korea}
\date{\today}
\maketitle

A recent paper by K. Xu {\it et al.} \cite{xu}
discussed a quantum depletion in gaseous
Bose-Einstein condensation.
In the ground state of an interacting Bose gas,
not all the particles are in the lowest energy state,
because the two-body interaction mixes the ground state
components with atoms in the other states.
For a homogeneous system the particle depletion is
known well from Bogoliubov theory as \cite{huan}
$ \eta_0 =\frac{8}{3 \sqrt{\pi}} \sqrt{\rho a^3},$
where $\rho$ is the atomic density
and $a$ is the s-wave scattering length.
The authors put the data of $^{23}$Na into above relation
and  say that ``For $^{23}$Na at a typical density
of $10^{14} cm^{-3}$, the quantum depletion is $0.2\%$".
Apparently, this is not a reliable value for the
inhomogeneous system which a magneto-optical trap is applied.
For an interacting and inhomogeneous system
the difficulty comes from the fact that
we do not know the every energy eigenvalue of the system
except $\vep_0$ which is obtained by
a macroscopic mean field wave function
so called Gross-Piataevskii equation(GPE).
However, one of the authors, W. Ketterle,  already discussed
the problem in an inhomogeneous system seriously
and obtained an effective method for that \cite{kett}.
We'll apply this method to obatin a better prediction.

We first consider a 3D system of $N$ non-interacting Bosons
confined by a harmonic potential of angular frequency $\omega$.
When the Bose distribution function in the grand canonical ensemble
at temperature $T$ is $f_B(\vep)$ and the density of states
of the isotropic 3D harmonic oscillator is $\rho(\vep)$,
the number of particles at temperature $T$  is obtained
by $N=\int f_B(\vep) \rho(\vep) d\vep$ \cite{kett,gros,haug,mull}.
It is
\begin{equation}
N=N_0+\left( \frac{k_B T}{\hbar\omega}\right)^3 g_3(z)
+\frac{2}{3}\left( \frac{k_B T}{\hbar\omega}\right)^2 g_2(z)
+ ...,
\label{20}
\end{equation}
where $g_n(z)=\sum_{l=1}^{\infty} z^l/l^n$
is the Bose functions.
 $z=e^{\beta(\mu-\vep_0)}$ is the fugacity between 0 and 1,
  $\vep_0 = (3/2)\hbar\omega$
 is the zero-point energy of the harmonic oscillator,
and $\mu(T)$ is the chemical potential.
$N_0$ is the ground state occupation number given by
$N_0=1/(z^{-1}-1)$.
The others are common notations.

For interacting system there is a shift of
the zero point energy by the interaction.
Then, the new ground state occupation number
is written as
\begin{equation}
N_0^{int} = \frac{1}{e^{\beta(\vep_0 +U -\mu)}-1},
\label{30}
\end{equation}
where $U=U(a,N_0)$ is the interaction energy between atoms
and may obtained from the solution of the GPE.
We consider the positive scattering length only,
so does $ U > 0 $.
Then, the condensate fraction of the interacting system
is definitely lower than that of the ideal system.

The chemical potential $\mu(T)$ is obtained numerically
by the following way.
The transition temperature $T_c$ is determined that at the
onset of condensation
$N_0 \rightarrow 0$ and $z \rightarrow 1$
from Eq. (\ref{20}). It is
\begin{equation}
T_c = T_0 \left\{1-\frac{\zeta(2)}{2\zeta(3)^{2/3}}
\frac{1}{N^{1/3}} + ... \right\} ,
\label{40}
\end{equation}
where
$T_0 = (\hbar\omega/k_B) \left\{ N/\zeta(3) \right\}^{1/3}$
and $\zeta$ is the Riemann-zeta function.
$T_c = 328 \sim 713 nK$ for $N=10^5 \sim 10^6$
 and $\omega = {10}^3 sec ^{-1}$.
The condensate fraction is obtained from Eq. (\ref{20})
as a function of temperature \cite{kett,gros,haug,mull}
\begin{equation}
\frac{N_0}{N}= 1-\left(\frac{T}{T_0}\right)^3
-\frac{3 \zeta(2)}{2 \zeta(3)^{2/3} N^{1/3}}
\left(\frac{T}{T_0}\right)^2+... .
\label{60}
\end{equation}
The author's prediction of the quantum depletion $0.2 \%$
provides that the transition temperature should be below
$40 \sim 90 nK$ for $N=10^5 \sim 10^6$.
Actually, the author's estimated value $\eta_0=0.2 \%$
 is too small for the inhomogeneous system to be accepted.
We obtain the fugacity $z(T)$ first
by solving Eq. (\ref{60}) numerically, and
then obtain the chemical potential $\mu(T)$.
We use this $\mu(T)$ for the estimation of $N_0^{int}$.

Comparing $N_0$ and $N_0^{int}$,
the over all behavior of the condensate fraction
in Eq. (\ref{30})
as a function of scattering length and temperature
can be estimated by the approximation
\begin{equation}
N_0^{int}(a,T) \sim N_{0}(T) e^{-\beta U(a,N_0)}.
\label{100}
\end{equation}
The ground state occupation number falls
exponentially as the interaction increases.


\end{document}